\documentclass[fleqn,11pt]{article}

\usepackage{graphicx}
\usepackage{amsmath,amssymb}
\usepackage{cite}
\usepackage{color}

\def\noi{\noindent}

\newcommand{\Title}[1]{\noi {{\Large\bf #1}}\\[1ex]}

\newcommand{\Author}[2]{\noi{\bf #1}\\[2ex]\noi{\normalsize\it #2}\\}

\newcommand{\Abstract}[1]{\vskip 2mm \begin{center}
        \parbox{16.4cm}{\small\noi #1} \end{center}\medskip}
\newcommand{\foom}[1]{\protect\footnotemark[#1]}

\def\email#1#2{\footnotetext[#1]{e-mail: #2}\addtocounter{footnote}{1}}

\def\nqq{\hspace*{-2em}}
\def\nhq{\hspace*{-0.5em}}

\def\cm{\hspace*{1cm}}
\def\inch{\hspace*{1in}}

\def\Jl#1#2{#1 {\bf #2},\ }

\def\ApJ#1 {\Jl{Astroph. J.}{#1}}
\def\CQG#1 {\Jl{Class. Quantum Grav.}{#1}}
\def\DAN#1 {\Jl{Dokl. AN SSSR}{#1}}
\def\GC#1 {\Jl{Grav. Cosmol.}{#1}}
\def\GRG#1 {\Jl{Gen. Rel. Grav.}{#1}}
\def\JETF#1 {\Jl{Zh. Eksp. Teor. Fiz.}{#1}}
\def\JETP#1 {\Jl{Sov. Phys. JETP}{#1}}
\def\JHEP#1 {\Jl{JHEP}{#1}}
\def\JMP#1 {\Jl{J. Math. Phys.}{#1}}
\def\NPB#1 {\Jl{Nucl. Phys. B}{#1}}
\def\NP#1 {\Jl{Nucl. Phys.}{#1}}
\def\PLA#1 {\Jl{Phys. Lett. A}{#1}}
\def\PLB#1 {\Jl{Phys. Lett. B}{#1}}
\def\PRD#1 {\Jl{Phys. Rev. D}{#1}}
\def\PRL#1 {\Jl{Phys. Rev. Lett.}{#1}}

\def\al{&\nhq}
\def\lal{&&{}\nqq}
\def\eq{Eq.\,}
\def\eqs{Eqs.\,}
\def\beq{\begin{equation}}
\def\eeq{\end{equation}}
\def\bear{\begin{eqnarray}}
\def\bearr{\begin{eqnarray} \lal}
\def\ear{\end{eqnarray}}
\def\earn{\nonumber \end{eqnarray}}
\def\nn{\nonumber\\ {}}

\def\nnn{\nonumber\\ \lal }
\def\nnnv{\nonumber\\[5pt] \lal }
\def\yy{\\[5pt] {}}
\def\yyy{\\[5pt] \lal }
\def\eql{\al =\al}

\def\sequ#1{\setcounter{equation}{#1}}

\def\dst{\displaystyle}
\def\tst{\textstyle}
\def\fracd#1#2{{\dst\frac{#1}{#2}}}
\def\fract#1#2{{\tst\frac{#1}{#2}}}
\def\Half{{\fracd{1}{2}}}
\def\half{{\fract{1}{2}}}

\def\e{{\,\rm e}}

\def\diag{\mathop{\rm diag}\nolimits}

\def\const{{\rm const}}

\def\eqn#1{\eq\eqref{#1}}
\def\rf{\eqref}
\def\mn{_{\mu\nu}}

\def\mN{_\mu^\nu}

\def\M{{\mathbb M}}

\def\kappa{\varkappa}

\def\GR{general relativity}
\def\cy{cylindrical}
\def\cyl{cylindrically symmetric}

\def\wh{wormhole}
\def\whs{wormholes}
\def\asflat{asymptotically flat}

\begin{document}
\twocolumn[
\thispagestyle{empty}

\Title{Potentially observable cylindrical wormholes\yy 
           without exotic matter in general relativity}

\Author{K. A. Bronnikov\foom 1}
	{\small Center for Gravitation and Fundamental Metrology, VNIIMS,
	   	Ozyornaya 46, Moscow 119361, Russia;\\
	Institute of Gravitation and Cosmology, Peoples' Friendship University of Russia
	(RUDN University),\\ \cm
		ul. Miklukho-Maklaya 6, Moscow 117198, Russia;\\
	National Research Nuclear University ``MEPhI''
		(Moscow Engineering Physics Institute),\\ \cm
		Kashirskoe sh. 31, Moscow 115409, Russia}

\Author{V. G. Krechet}
	{\small Moscow State Technological University ``Stankin'',
         Vadkovsky per. 3A, Moscow 127055, Russia} 

\Abstract
{All known solutions to the Einstein equations describing rotating
 cylindrical wormholes lack asymptotic flatness in the radial directions and therefore cannot
 describe wormhole entrances as local objects in our Universe. To overcome
 this difficulty, wormhole solutions are joined to flat asymptotic regions 
 at some surfaces $\Sigma_-$ and $\Sigma_+$. The whole configuration
 thus consists of three regions, the internal one containing a wormhole throat,
 and two flat external ones, considered in rotating reference frames. Using a
 special kind of anisotropic fluid respecting the Weak Energy Condition (WEC)
 as a source of gravity in the internal region, we show that the parameters 
 of this configuration can be chosen in such a way that matter on both junction
 surfaces $\Sigma_-$ and $\Sigma_+$ also respects the WEC. Closed timelike 
 curves are shown to be absent by construction in the whole configuration.  It seems 
 to be the first example of regular twice (radially) asymptotically flat wormholes without 
 exotic matter and without closed timelike curves, obtained in general relativity.
}

\medskip
] 
\email 1 {kb20@yandex.ru}
\section{Introduction}

  Traversable Lorentzian wormholes are widely discussed in gravitational
  physics since they lead to many effects of interest like time machines
  or shortcuts between distant parts of space. Large enough \whs, if any,
  can lead to observable effects in astronomy \cite{sha, accr, kir-sa1, kb-bal-19}.

  In attempts to build realistic \wh\ models, the main difficulty is that
  in \GR\ (GR) and some of its extensions a static \wh\ geometry requires
  the presence of ``exotic'', or phantom matter, that is, matter violating
  the weak and null energy condition (WEC and NEC), at least near the
  throat, the narrowest place in a \wh\ \cite{thorne,viss-book,HV97,ws_book}.
  These results were obtained if the throat is a compact 2D surface with a
  finite area \cite{HV97}. Assuming asymptotic flatness and fulfillment of 
  the averaged NEC, topological restrictions have been proven
  \cite{frie93, galloway, frie06} that forbid the existence of \whs\ having
  two flat asymptotic regions (the so-called topological censorship).  

  Examples of phantom-free \wh\ solutions are known in extensions
  of GR, such as the Einstein-Cartan theory \cite{BGal15,BGal16},
  Einstein-Gauss-Bonnet gravity \cite{GBo}, brane worlds \cite{BKim03}
  and other multidimensional models \cite{BS16}, etc.
  We here prefer to adhere to GR as a theory well describing the
  macroscopic reality while the extensions more likely concern
  very large densities and/or curvatures. In GR there are phantom-free \wh\
  models with axial symmetry, such as the Zipoy \cite{zipoy} and
  superextremal Kerr vacuum solutions as well as solutions with 
  scalar and electromagnetic fields \cite{br-fab97, matos15}; 
  in all of them, however, a disk that plays the role of a throat is bounded 
  by a ring singularity whose existence is a kind of unpleasant price paid for the 
  absence of exotic matter. Regular phantom-free \whs\ in GR were found in 
  \cite{canf1,canf2}, sourced by a nonlinear sigma model, but they are asymptotically 
  NUT-AdS instead of the desired flatness. A phantom-free \wh\ construction in 
  \cite{schhein} contains singularities and closed timelike curves. These
  shortcomings may be interpreted as manifestations of topological censorship. 

  The above-mentioned results of \cite{HV97} as well as topological censorship
  are not directly applicable to objects like cosmic strings, infinitely stretched along a certain 
  direction, in the simplest case \cyl\ ones. Thus, for example, nontrivial stationary \cyl\ 
  systems cannot be completely \asflat\ since in the longitudinal ($z$) direction, due to 
  $z$-inde\-pen\-dence, at large $z$ the curvature is the same as at small $z$, and which is 
  important, such nonzero curvature is preserved along null ($z,t$) directions owing to time 
  independence. In other words, all nontrivial \cyl\ systems are not asymptotically flat in the
  usual sense. And it is this circumstance that gives us a hope to obtain a \wh\ without 
  exotic matter that will  be \asflat\  in the remaining two spatial directions (or, which is 
  the same, in the radial direction) on both sides of the throat (maybe up to an angular 
  deficit, as in cosmic strings), which is necessary if we wish it to be potentially visible to
  distant observers like ourselves residing in weakly curved  regions of the Universe. 
  
  We can remark that cylindrical symmetry was used for many decades as a 
  kind of theoretical laboratory, where one could ask, for example, what can happen 
  under extremely large deviations from spherical symmetry, or study anisotropic
  cosmological models. The corresponding isometry group provides many mathematical 
  results of interest (see, e.g., \cite{exact-book}
  and references therein). On the other hand, the fields of some natural objects (jets, 
  filament-like structures etc.) may be approximately described as \cyl\ ones in some 
  restricted region. However, studies with this symmetry have gained much more interest 
  and popularity since the theoretical discovery of cosmic strings, leading to attempts 
  to find them in the Universe and to use them for solving a number of astrophysical and 
  cosmological problems \cite{vilenkin}. Cylindrical \whs, if any, may look like cosmic 
  strings for a distant observer.
  
  Cylindrical \whs\ with and without rotation were discussed, in particular, in  
  \cite{BLem09,BLem13,BS14,kb-conf16,BK15} (see also references therein). 
  It was shown, with a number of examples, that phantom-free \cy\ \wh\ solutions 
  to the Einstein equations are easily obtained.  A problem with \cy\ systems is, as in 
  \cite{canf1,canf2}, their undesirable asymptotic  behavior. This does 
  not look unexpected since even in Newtonian theory the gravitational potential 
  of a cylindrical body grows logarithmically at large radii, and its relativistic counterpart,
  the Levi-Civita vacuum solution \cite{LC}, has similar properties. 
  The majority of papers devoted to \cy\ systems in gravity theories do not care of a possible 
  even partial asymptotic flatness, frequently discussing  matter distributions matched to the 
  Levi-Civita (or Lewis \cite{lewis}) external solutions. Our study (as well as 
  \cite{BLem13,BK15,kb-conf16}) has the advantage that the requirement of radial 
  asymptotic flatness is our basic concern, and it is a maximum of what could be required 
  under cylindrical symmetry. 
    
  To overcome this difficulty with \cy\ \wh\ solutions and to provide radial asymptotic flatness, 
  it was suggested \cite{BLem13} to cut such a solution on some surfaces (cylinders)
  $\Sigma_-$  and $\Sigma_+$ on both sides of the throat and to join them to suitable 
  parts of Minkowski space-time, $\M_-$ and $\M_+$. 
  Each of the latter should have a spatial part in the form of Euclidean space with a 
  cut-out straight tube of finite radius. The surfaces $\Sigma_-$ and $\Sigma_+$ then 
  contain some matter whose stress-energy tensor (SET) components $S_a^b$ 
  are determined by the junction conditions in terms of jumps of the extrinsic
  curvature \cite{israel-67, BKT-87}.  A \wh\ model without exotic matter is thus built if 
  matter in the internal region and the surface matter on both $\Sigma_-$ and 
  $\Sigma_+$ respect the WEC. No successful examples of phantom-free \whs\ were 
  so far obtained in this way. Moreover, it was shown that many kinds of matter 
  filling the internal region create such geometry that it is impossible to obtain 
  $S_a^b$ satisfying the WEC on both junctions $\Sigma_\pm$ \cite{BK15,kb-conf16}.

  In the present paper we show that this goal is achieved if we use a special kind
  of anisotropic fluid as a source of gravity in the internal region.  In the next section we
  obtain the internal solution, in Section 3 we consider its matching to flat external regions
  and show that the whole model satisfies the WEC under a proper choice of the free
  parameters. Section 4 contains a discussion of a difficulty emerging due to 
  different signs of the angular velocity of rotation $\Omega$ in $\M_+$ and $\M_-$. 
  The problem emerges if we try to replace a thin shell with a smooth matter distribution: 
  in the latter, described in its comoving reference frame, the rotational direction cannot 
  change from one layer to another. A suggested way out is to use the fact that in vacuum 
  all reference frames are comoving. The Appendix contains a calculation related to this 
  discussion: it is shown that using the presently studied wormhole solution, it is 
  impossible to obtain the same sign of $\Omega$ in both $\Sigma_+$ and $\Sigma_-$.  

\section{Wormhole solution\\ with an anisotropic fluid}

  Consider a stationary \cyl\ metric
\bearr                                                    \label{ds-rot}
         ds^2 = \e^{2\gamma(x)}[ dt - E(x)\e^{-2\gamma(x)}\, d\varphi ]^2
       - \e^{2\alpha(x)}dx^2 
\nnn \cm
	- \e^{2\mu(x)}dz^2 - \e^{2\beta(x)}d\varphi^2,
\ear
  where $x$, $z$ and $\varphi$ are the radial, longitudinal and angular
  coordinates. This metric is said to describe a \wh\ if either (i) the
  circular radius $r(x) = \e^{\beta(x)}$ has a regular minimum (called an
  $r$-throat) and is large or infinite far from this minimum or (ii) the 
  same is true for the area function $a(x) = \e^{\mu+\beta}$ 
  (its minimum is called an $a$-throat) \cite{BLem09, BLem13}. If a \wh\ 
  is \asflat\ at both extremes of the $x$ range, it evidently possesses both 
  kinds of throats.

  The metric coefficient $g_{03} = -E$ corresponds to space-time rotation
  which can be characterized by the angular velocity $\omega(x)$ of a congruence 
  of timelike curves \cite{BLem13, kr2, kr4},
\beq                                                  \label{om}
          \omega = \half (E\e^{-2\gamma})' \e^{\gamma-\beta-\alpha}.
\eeq
  (this expression holds under an arbitrary choice of the coordinate $x$, and
  a prime stands for $d/dx$). Furthermore, in the reference frame comoving 
  to matter in its motion by the angle $\varphi$ we have the SET component
  $T^3_0 = 0$, hence (via the Einstein equations) we have the Ricci tensor 
  component $R_0^3 \sim (\omega \e^{2\gamma+\mu})' = 0$, so that
  \cite{BLem13} 
\beq       	      					\label{omega}
	\omega = \omega_0 \e^{-\mu-2\gamma}, \cm \omega_0 = \const.
\eeq
  Then, according to \rf{om}, 
\beq                          \label{E}
	E(x) = 2\omega_0 \e^{2\gamma(x)} \int \e^{\alpha+\beta-\mu-3\gamma}dx.
\eeq

  It then turns out \cite{BLem13} that the diagonal components of the Ricci
  ($R\mN$) and Einstein ($G\mN = R\mN - \half \delta\mN R$) tensors split
  into those for the static metric (that is, (\ref{ds-rot}) with $E=0$) plus an 
  $\omega$-dependent addition:
\bear              \label{split}
		R\mN \eql {}_s R\mN + {}_\omega R\mN, \qquad
		G\mN = {}_s G\mN + {}_\omega G\mN, 
\nn		
	{}_\omega R\mN \eql \omega^2 \diag (-2, 2, 0, 2),  
\nn      
   	{}_\omega G\mN \eql \omega^2 \diag (-3, 1, -1, 1),  
\ear
  where ${}_s R\mN$ and ${}_s G\mN$ are the static parts.  
  The tensors ${}_s G\mN$ and ${}_\omega G\mN$ (each separately)
  satisfy the conservation law $\nabla_\alpha G^\alpha_\mu =0$ in terms of this
  static metric. Thus, by the Einstein equations $G\mN = - \kappa T\mN$ 
  ($\kappa = 8\pi G$), the tensor ${}_\omega G\mN/\kappa$ acts as an 
  additional SET with exotic properties (e.g., the effective energy density is 
  $-3\omega^2/\kappa <  0$), making it easier to obtain both $r$- and 
  $a$-throats, as confirmed by a number of examples in \cite{kr4,BLem13,BK15}.

  Such \whs, however, cannot be \asflat\ since the latter would require 
  $\omega \to 0$ along with finite limits of $\gamma$ and $\mu$, which is
  incompatible with (\ref{omega}).

  To obtain radially \asflat\ models, it was suggested \cite{BLem13} to cut our 
  \wh\ solution at some regular cylinders $\Sigma_+\,(x=x_+)$ and 
  $\Sigma_-\,(x=x_-)$ on both sides of the throat and to join it 
  there to flat-space regions extending to infinity. Such junction surfaces
  comprise thin shells with certain surface SETs, and it remains to check 
  whether these SETs satisfy the WEC and NEC.

  It turns out \cite{BK15, kb-conf16} that with many kinds of matter sources 
  of the \wh\ solutions it is impossible to obtain surface SETs respecting 
  the NEC on both $\Sigma_+$ and $\Sigma_-$. This happens if 
  $T^t_t = T^\varphi_\varphi$, which holds, e.g., for scalar fields with 
  arbitrary self-interaction potentials and for an azimuthal magnetic field
   ($F_{21}=-F_{12}\ne 0$, where $F\mn$ is the Maxwell tensor). So, even 
  without solving the field equations, we can be sure that the solution is 
  not suitable for making a twice \asflat\ \wh\ free from exotic matter.  

  Let us, instead, consider an anisotropic fluid that respects the WEC,
  in its comoving reference frame (the 4-velocity is $u^\mu = \delta^{\mu 0} \e^{-\gamma}$),
  with a SET having the nonzero components\footnote
    {This SET is chosen by analogy with the SET of a longitudinal magnetic field 
    in a static \cyl\ space-time, which cannot be directly extended to 
    models with rotation.}
\bearr   \label{SET}
	   T^0_0 = - T^1_1 = T^2_2 = - T^3_3 = \rho(x),
\nnn 
	   T^0_3 = -2\rho E \e^{-2\gamma},
\yyy        \label{rho}
       \rho = \rho_0 \e^{-2\gamma - 2\mu},  \quad\ \rho_0 = \const > 0,
\ear 
  where \eqn{rho} follows from the conservation equation $\nabla_\mu T^\mu_1 =0$
  (for a full presentation of the anisotropic fluid formalism in stationary \cy\
  space-times see, e.g., \cite{santos}). 
  
  That the WEC is really fulfilled for the SET \rf{SET}, can be verified by finding the principal 
  pressures $p_i$ as the eigenvalues of the tensor $T\mN$ written in orthonormal tetrad 
  components, $T_{(mn)} = e^\mu_{(m)} e^\nu_{(n)} T\mn$, where the parentheses mark 
  tetrad indices ranging from 0 to 3. The WEC requires 
\beq            \label{WEC}                     
	    \rho \geq 0, \quad\ \rho + p_i \geq 0.
\eeq  
  Choosing the tetrad
\bearr            \label{tetrad}
		e^\mu_{(0)} =  (\e^{-\gamma} , 0,0,0), \ \ 
		e^\mu_{(1)} =  (0,\e^{-\alpha} , 0,0),
\nnn 
		e^\mu_{(2)} =  (0,0,\e^{-\mu}, 0),\ \ 
		e^\mu_{(3)} =  (E \e^{-\beta - 2\gamma} , 0,0,e^{-\beta}),		
\nnn		
\ear
  it is straightforward to obtain that the principal pressures for the SET \rf{SET} are
\beq
                p_x = \rho, \quad\ p_z = - \rho, \quad  p_\varphi = \rho,
\eeq    
  and the conditions \rf{WEC} are satisfied.
  
  To solve the Einstein equations, it is sufficient to consider the diagonal components, 
  their single off-diagonal component then automatically holds as well \cite{BLem13}
  since $G^0_3 = E \e^{-2\gamma} (G^3_3-G^0_0)$, and a similar relation holds for $T\mN$
  components.  In terms of the harmonic radial coordinate $x$, such that 
\beq             \label{harm}
  	\alpha = \beta+\gamma+\mu,
\eeq  
  the diagonal components of the Einstein equations read
\bear           \label{R00}
  	\e^{-2\alpha} \gamma'' + 2\omega^2 \eql \kappa\rho,
\yy		\label{R22}
	\e^{-2\alpha} \mu'' \eql \kappa\rho,
\yy		\label{R33}
	\e^{-2\alpha} \beta'' - 2\omega^2 \eql - \kappa\rho,
\yy			\label{G11}
	\e^{-2\alpha} (\beta'\gamma' + \beta'\mu' + \gamma'\mu') + \omega^2 \eql \kappa\rho.
\ear
  A sum of \rf{R00} and \rf{R33} gives $\beta''+\gamma''=0$, whence
  $\e^\beta = r_0 \e^{-\gamma + \gamma_1 x}$, where $r_0$ (to be used as a length scale) 
  and $\gamma_1$ are constants, and we put $\gamma_1 =0$ for simplicity. 
  This removes $\mu'$ from \eq \rf{G11}, and its integration gives
\bearr                                       \label{L1}
      r^2 \equiv \e^{2\beta}= \frac{r_0^2}{Q^2 (x_0^2-x^2)}, \quad\ 
		\e^{2\gamma} = Q^2 (x_0^2-x^2), 
\nnn 
		x_0 := \frac{\omega_0}{\kappa\rho_0 r_0}, \qquad\ Q^2 := \kappa \rho_0 r_0^2.      
\ear
  The constants $x_0$ and $Q$ thus defined are dimensionless, while $r$ and $\e^{\alpha}$ 
  have the dimension of length. Next, $\e^\mu$ is obtained by integrating \rf{R22}:
\beq 						\label{L2}
		\e^{2\mu} = \e^{2mx} (x_0 -x)^{1-x/x_0} (x_0 +x)^{1+ x/x_0},		
\eeq
  where $m = \const$, and one more constant has been suppressed by rescaling the $z$ axis.
  Lastly, $E(x)$ is found using \rf{E} with \rf{harm}, \rf{L1}, \rf{L2}:
\beq      \nhq         \label{L3}
        E = \frac {r_0(x_0^2 - x^2)} {2x_0^2} \bigg[ \frac {2x_0 x}{x_0^2 - x^2} 
        		+  \ln\frac{x_0+x}{x_0 -x} + E_0\bigg], 	          
\eeq
  where $E_0 = \const$.

  The solution \rf{rho}, \rf{L1}, \rf{L2}, \rf{L3} contains five integration constants 
  $\omega_0$, $\rho_0$, $r_0$, $m$, and $E_0$, and the coordinate $x$ ranges from 
  $-x_0$ to $x_0$. The circular radius $r \to \infty$ as $x\to \pm x_0$, thus confirming a
  \wh\ nature of the geometry, but in the same limits $\e^\gamma \to 0$, so $x = \pm x_0$ 
  are curvature singularities, at which the Kretschmann scalar behaves as $|x_0 - x|^{-4}$.  
  
  A question of interest is whether the space-time described by this solution contains 
  closed timelike curves (CTCs) that lead to causality violation. In the metric \rf{ds-rot}
  such curves emerge if and only if $g_{33} > 0$ (see, e.g., \cite{gron}), in which case the 
  closed coordinate lines of the azimuthal angle $\varphi$ are timelike. Consider the behavior
  of $g_{33}$ for the symmetric branch of the above solution, to be used in what follows. 
  This branch corresponds to $m = E_0 =0$, implying that all metric coefficients are even 
  functions of $x$, except for $E$ which is odd. For this symmetric solution, 
\beq  \nhq                   \label{CTC}
	      g_{33} = \frac{r_0^2}{{x_0^2 - x^2}} 
	      	\bigg[ -1 + \bigg(y + \frac {1 - y^2}{2} \ln \frac{1 + y}{1 - y}\bigg)^2\bigg],
\eeq    
  where $y = x/x_0$. An inspection shows that $g_{33} > 0$ and hence there are CTCs at 
  $|y| > 0.564$, i.e., close enough to the singularities that occur at $y = \pm1$.

 \section{Potentially observable models}

  Let us now try to construct a twice (radially) \asflat\ \wh\ configuration. To do that, we take the \wh\ 
  metric described above, cut it at some regular points $x=x_-$ to the left and $x= x_+$ to the 
  right of its throat $x=0$ and join it at $x=x_\pm$ to regions $\M_{\pm}$ of Minkowski
  space-time with the metric $ds_{\rm M}^2 = dt^2 - dX^2 - dz^2 - X^2 d\varphi^2$ 
  taken in \cy\ coordinates (at some $X = X_\pm = \const$). To be able to match it to 
  \rf{ds-rot} with $E \ne 0$, we transform it to a rotating reference frame by substituting  
  $\varphi \to \varphi + \Omega t$,  $\Omega = \const$, whence it follows
\beq                                                          \label{ds_M}
	      ds_{\rm M}^2 = dt^2 - dX^2 - dz^2 - X^2 (d\varphi + \Omega dt)^2.
\eeq
  Then the relevant quantities in terms of (\ref{ds-rot}) are
\bearr                                                   \label{M-param}
      \e^{2\gamma} =  1 - \Omega^2 X^2,\qquad
      \e^{2\beta} = \frac{X^2}{1 - \Omega^2 X^2},
\nnn
      E = \Omega X^2, \qquad    \omega = \frac{\Omega}{1 - \Omega^2 X^2}.
\ear
  This metric is stationary and ready for matching to an internal metric at $|X| <  1/|\Omega|$, 
  so that the linear rotational velocity is smaller than the velocity of light.

  Matching at a surface $\Sigma:\ x=x_*$ means that we identify the two 
  metrics on this surface, so that
\beq                                                       \label{ju-1}
      [\beta] = 0, \quad [\mu] = 0, \quad     [\gamma] = 0, \quad [E] =0,
\eeq
  with the conventional notation for discontinuities: $[f] = f(x_*+0) - f(x_* -0)$ for any $f(x)$. 
  Then the coordinates $t, z, \phi$ can be identified in the whole space. However, the choice 
  of radial coordinates may differ on different sides of the junction surface, and it is admissible 
  since all quantities used in the matching conditions are insensitive to the choice of $x$ or $X$.  

  At the next step, we should calculate the SET of matter on the junction surface using
  the Darmois-Israel formalism  \cite{israel-67,BKT-87}. In our case of a timelike surface
  $x = x^1 = \const$, the SET $S_a^b$ is expressed in terms of the extrinsic curvature 
  $K_a^b$ as
\def\tK {{\tilde K}{}}
\bearr                                                        \label{ju-2}
	S_a^b =  \kappa^{-1} [\tK_a^b], \qquad
		\tK_a^b := K_a^b - \delta_a^b K^c_c, 		
\ear
 where the indices $a, b, c = 0, 2, 3$, and $K_{ab} = \half \e^{-\alpha} g'_{ab}$, the prime 
 denotes a derivative with respect to $x$ in the internal region and with respect to $X$ in 
 $\M_\pm$.  In the notations of \rf{ds-rot}, the nonzero components of $\tK_{ab}$ are
 \bearr                              \label{tKab}
      \tK_{00} = - \e^{-\alpha+2\gamma} (\beta' + \mu'),    
\nnn      
      \tK_{03} = -\Half \e^{-\alpha} E' + E\e^{-\alpha}(\beta'+\gamma'+\mu'),
\nnn
	\tK_{22} = \e^{-\alpha + 2\mu} (\beta'+\gamma'),
\nnn 
	\tK_{33} = \e^{-\alpha+2\beta}(\gamma'+\mu') 
\nnn \qquad
	+ \e^{-\alpha-2\gamma}[EE'- E^2(\beta'+2\gamma'+\mu')]. 	      
\ear      
   
 We wish to find out whether the surface SETs on both $\Sigma_\pm$ satisfy the WEC 
 requirements 
\beq                                                             \label{WEC-s}
	S_{00}/g_{00} = \sigma \geq 0, \qquad\
		S_{ab}\xi^a \xi^b \geq 0,
\eeq
  where $\xi^a$ is any null vector ($\xi^a \xi_a =0$) on $\Sigma=\Sigma_\pm$.
  The second inequality in (\ref{WEC-s}) comprises the NEC as part of the WEC.
  The conditions (\ref{WEC-s}) are equivalent to
\beq                                                            \label{WEC1}
      [\tK_{00}/g_{00}] \geq 0, \qquad\ [K_{ab}\xi^a \xi^b] \geq 0.
\eeq
  
  Consider the matching conditions (\ref{ju-1}) on $\Sigma_\pm$, identifying
  the surfaces $X=X_\pm$ in Minkowski regions and $x=x_\pm$ in the internal region. 
  The conditions $[\beta] = [\gamma] =0$ at any of the two junctions lead to
\bearr  					\label{ju-3}
		r_0^2[Q^2 (x_0^2 - x^2)]^{-1} = X^2 [1-\Omega^2 X^2]^{-1},
\yyy  					\label{ju-4}
		Q^2 (x_0^2 - x^2) = 1-\Omega^2 X^2 = :P,
\ear
  where, without risk of confusion, we have omitted the index $\pm$ at $x$ and $X$.
  From \rf{ju-3} and \rf{ju-4} it follows $X_\pm = \pm r_0$. Thus the value of $X$ suitable for 
  matching with our \wh\ solution is fixed by the length scale $r_0$. We will also assume 
  $x_- = -x_+$, and then due to \rf{ju-4} $\Omega^2$ is the same in $\M_+$ and $\M_-$.  
  Next, the condition $[\mu]=0$ is easily achieved by choosing a $z$ scale in $\M_+$ and 
  $\M_-$ and does not lead to any restrictions. Lastly, the condition $[E] =0$ gives
\beq                                        \label{ju-5}
  	\pm 2x_0^2 \sqrt{1-P} = 2xx_0 + (x_0^2 -x^2) \ln \frac {x_0+x}{x_0-x},
\eeq  	
  In the derivation of \rf{ju-5} we have used the assumption $E_0 =0$ in \eqn{L3}, so that
  $E(x)$ is an odd function, and $E (x_+) = - E(x_-) \ne 0$. Since in the Minkowski
  regions $E = \Omega X^2$ (see \rf{M-param}) while $\Omega^2 = (1- P)/X^2$, 
  equal values on both junctions, we have to conclude that $\Omega(\M_-) = - \Omega(\M_+)$. 
  Therefore on $\Sigma_\pm$ we have $E = \pm \sqrt{\Omega^2} X^2 = \pm r_0 \sqrt{1-P}$. 
  Comparing this with \rf{L3}, we arrive at \rf{ju-5}.     
      		
  Now let us try to choose such values of the free parameters of the solution 
  that the conditions \rf{WEC-s} will be satisfied. The condition $\sigma \geq 0$ 
  leads to
\beq               \label{sigma+}
            [\e^{-\alpha+2\gamma} (\beta' + \mu')] \leq 0.   
\eeq  
   However, the second condition (the NEC) is not so easily formulated since it should hold for 
   any null vector in $\Sigma_\pm$. One could analyze it directly, using a one-parameter 
   family of vectors $\xi^a$ representing all possible null directions on $\Sigma_\pm$
   This will lead to rather cumbersome expressions, not too easy to handle.  
   
   Instead, we recall that the WEC fulfillment (including the NEC) may be verified by finding 
   the density and  principal pressures in a comoving reference frame and applying the 
   conditions \rf{WEC} to our surface quantities $\sigma$ and $p_i$.  A certain difficulty is 
   that in our case both $\Sigma_\pm$ are not considered in comoving frames (the latter
   would require $[\omega] = 0$). 
  
   Still it is not necessary to find an explicit transformation to the comoving frame for 
   the surface matter, which is not so easy. Instead, we can find the values of $\sigma$ 
   and the principal pressures $p_z$, $p_\varphi$ in this frame as eigenvalues of 
   the surface SET represented in a local Minkowski (tangent) space, thus avoiding any 
   distortions due to curvature or coordinate choice. To do that, we can use an orthonormal
   triad formed by three of the four vectors \rf{tetrad}, excluding $e^\mu_{(1)}$, the one  
   orthogonal to $\Sigma_\pm$, while others are tangent to it.
   
   Such a calculation leads to the following results: in  the orthonormal triad on $\Sigma_\pm$
\bearr 				\label{triad}
              e_{(0)}^a = (\e^{-\gamma},0,0);    \qquad
              e_{(2)}^a = (0, \e^{-\mu}, 0); 
\nnn
              e_{(3)}^a = (E\e^{-\beta-2\gamma}, 0, \e^{-\beta}).           
\ear    
   (where the parentheses mark triad indices), the discontinuities $[\tK_{(ab)}]$ form the matrix
\beq   				\label{matrix}                          
		[\tK_{(ab)}] =
		\begin{pmatrix}             a & 0 & d \\
						       0 & b & 0 \\
						       d  & 0 & c \end{pmatrix},
\eeq 
  where 
\bearr                 \label{abcd}
  	  a = -[\e^{-\alpha}(\beta'+\mu')],  \quad\ \,
  	  b = [\e^{-\alpha}(\beta' + \gamma')],
\nnn  	  
  	  c = [\e^{-\alpha}(\gamma'+ \mu')],\qquad\
  	  d = -[\omega]
\ear	     			
   The eigenvalues of the matrix \rf{matrix} are easily found as roots of its characteristic equation:
\beq   				\label{roots}                          
             \Big\{\half(a + c  \pm \sqrt{(a-c)^2 +4 d^2}),\  b \Big\}.
\eeq      

   The SET under consideration has the form $S_{(mn)} = \diag (\sigma, p_z, p_\varphi)$, 
   but it is not at once evident which of the eigenvalues \rf{roots} corresponds to a
   particular SET component. To make it clear, we notice that if the reference frame is 
   initially comoving, that is, if $d=0$, we have $(\sigma, p_z, p_\varphi) \propto (a,b,c)$.
   Accordingly, we can take, with a common proportionality factor,
\bearr         \label{roots1}
	(\sigma, p_z, p_\varphi) \propto 	\Big(a+c+ S,\  2b,\ a+c -S \Big),
\nnn
	S := \sqrt{(a-c)^2 +4 d^2}),
\ear
  assuming $a-c >0$ (otherwise $a$ and $c$ should be interchanged).
  As a result, the WEC requirements \rf{WEC1} read
\bearr        \label{Wa} 
	a+c+\sqrt{(a-c)^2 +4 d^2} \geq 0,
\yyy          \label{Wb}    
	a+c+\sqrt{(a-c)^2 +4 d^2} + 2b \geq 0,
\yyy            \label{Wc}
	a + c \geq 0.          
\ear                                 
\begin{figure*}  
\centering
          \includegraphics[width=5.8cm]{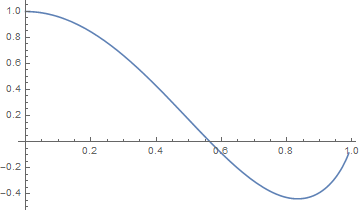}
          \includegraphics[width=5.8cm]{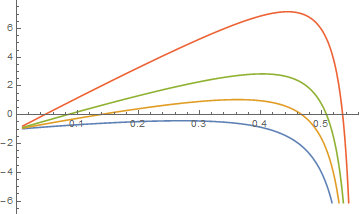}
          \includegraphics[width=5.8cm]{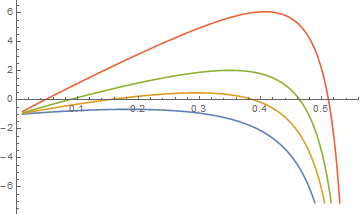}
\caption{\small
          Left: The function $P(y)$. The plot shows the range of acceptable values of $y$.
          Middle: $a(y)$ for $x_0 = 0.3,\ 0.4,\ 0.5,\ 0.75$ (upside down). 
          Right: $a(y)+c(y)$ for $x_0 = 0.3,\ 0.4,\ 0.5,\ 0.75$  (upside down).
          The data for $x_0 =0.75$ are included to show that larger $x_0$ values inevitably 
          lead to WEC violation.            
              }  
\end{figure*}  
  This reasoning, beginning with \rf{sigma+}, is of quite a general nature for any surfaces 
  $x=\const$ in space-times with the metric \rf{ds-rot}. Let us now specify the matrix elements 
  $a,b,c,d$ for the surfaces $\Sigma_\pm$ in our particular construction, 
  assuming for certainty $\omega_0 > 0$ in the internal solution and $\Omega > 0$ in $\M_+$.
    
  Consider  $\Sigma_+$, hence for any $f$ we write $[f] = f_{\rm out} - f_{\rm in}$,
  where $f_{\rm out}$ is taken from the region $\M_+$ with the metric \rf{ds_M} at $X= r_0$, 
  and $ f_{\rm in}$ from the \wh\ solution \rf{L1}--\rf{L3} at $x = x_+$. Taking into account the 
  junction conditions \rf{ju-3}--\rf{ju-5}, we obtain (ignoring the insignificant common factor $1/r_0$):
\bearr                  \label{sig-abc}
	a =  - \frac 1{P(y)} + \frac {M(y)}{x_0^2}\bigg(\frac{y}{1-y^2} + \Half L(y)\bigg),
\nnnv	
	b = 1,
\nnnv
        c= - \frac 1{P(y)} + 1 + \frac {M(y)}{x_0^2}\bigg(\frac{y}{1-y^2} - \Half L(y)\bigg),
\yyy				\label{sig-d}
	d = -\frac{\sqrt{1-P(y)}}{P(y)} + \frac{M(y)}{x_0^2(1-y^2)}, 
\ear      
   where we use the notations 
\bearr                       \label{not-flu}
                 y = \frac{x_+}{x_0}, \cm L(y) = \ln \frac{1+y}{1-y},
\nnnv 
                 M(y) = \big(1-y\big)^{-(1-y)/2}\big(1+y\big)^{-(1+y)/2},
\ear 
  and for the quantity $P$ defined in \rf{ju-4} we have 
\beq                 \label{not-P}
		P(y) = (1-y^2)\Big[1 - y L(y) - \frac 14 (1-y^2) L^2(y)\Big].
\eeq

  The expressions for $a,b,c$ depend on two parameters, $x_0$ and $y$, and, 
  by symmetry of our construction, they are the same on both $\Sigma_+$ and $\Sigma_-$. 
  Unlike that, when calculating $d = -[\omega]$ for $\Sigma_-$, we should take
   into account different signs of $\Omega$ on the two junctions. As a result, on $\Sigma_-$ 
   in the expression \rf{sig-d} for $[d]$ there are minuses at both terms (whose own values
   are the same as in \rf{sig-d}), and so the absolute value of $[d]$ is larger than on $\Sigma_+$. 
   However, in our WEC analysis, the particular value of  $d = - [\omega]$ is not important since, 
   as follows from \rf{Wa}--\rf{Wc}, the WEC fulfillment only depends on $a$ and $c$ if  
   $b \geq 0$, and in our case $b=1$. 
     
    In \eqs \rf{sig-abc}--\rf{not-P} all quantities are dimensionless, and one can 
    verify the following (see Fig.\,1):

\medskip\noi            
  $\bullet$ The condition $0< P(y) < 1$, necessary for all the above expressions to 
  make sense, holds for $0< |y| < 0.564)$ (this and further numerical estimates 
  are approximate). One can notice that the admissible range of $y$ is the same as 
  was previously found for the absence of CTCs. This is not surprising since $g_{33}$
  on $\Sigma_\pm$ is common for the internal and external regions, and the latter, being 
  flat, is manifestly free of CTCs. Since $|y|$ in the internal region is smaller than on 
  $\Sigma_\pm$, this region is also CTC-free.

\medskip\noi        
  $\bullet$ The condition $a > 0$ (equivalent to $\sigma >0$ in the noncomoving
   reference frame on $\Sigma_\pm$ in which the junction conditions were written) holds, 
   in particular, for $x_0 = 0.5,\ y \in (0.15, 0.47)$ and for $x_0 = 0.3,\ y \in (0.05, 0.53)$.

\medskip\noi
  $\bullet$  The condition $a > c$, necessary for interpreting the roots as shown 
  in \rf{roots1}, holds practically in the same ranges of $y$ for $x_0$ 
  (at least) between 0.3 and 0.5.  

  \medskip\noi  
  $\bullet$ The condition $a+c >0$ also holds in a sufficiently large range of the 
  parameters, e.g, for $x_0 = 0.5,\ y \in (0.15, 0.38)$ and for $x_0 = 0.3,\ y \in (0.05, 0.51)$.
  
\medskip
  Since the conditions \rf{Wa} and \rf{Wb} manifestly hold if \rf{Wc} does, we conclude 
  that there is a significant range in the parameter space ($x_0, y$) in which our \wh\ model 
  completely satisfies the WEC.
 
\section{Concluding remarks}

  Using an explicit example of an anisotropic fluid as a source of gravity, we have 
  achieved our goal and demonstrated the possibility of obtaining a regular, radially \asflat\ 
  traversable \cy\ \wh\ in GR.

  There is, however, a subtle point that may put to doubt the consistency of the whole 
  construction.\footnote
  	{We are grateful to the anonymous referee for pointing out this problem.}
  The present model contains thin shells on two surfaces and is consistent in the 
  framework of the thin shell formalism. However, if a thin shell is physically 
  understood as some approximation to a smooth thick layer with matter content 
  rapidly varying in a region of finite extent, there emerges a discrepancy with the 
  shell (here, $\Sigma_-$) that separates the internal region with $\omega_0 > 0$ 
  with the external one where $\Omega = \Omega_- < 0$. For a smooth, differentially 
  rotating matter in a thick shell, one can define a comoving frame in which 
  \eqn{omega} should hold everywhere with fixed $\omega_0 > 0$. It is then hard 
  to understand how it can be smoothly joined to the outside $\Omega < 0$. 
  
  A possible answer is to recall that in vacuum (hence in the region $\M_-$) any 
  reference frame is comoving. Therefore we can imagine that the layer of matter
  with $\omega > 0$ is smoothly matched to a reference frame in $\M_-$ with 
  $\Omega > 0$, but for the description of the whole configuration we are using 
  there a frame in which $\Omega = \Omega_- < 0$. Technically, this corresponds
  to a rapid change of $\Omega = \Omega(x)$ in a vacuum layer close to the matter
  distribution replacing the thin shell.\footnote
  	  {Note that it is in any case necessary to consider different reference frames in 
  	   $\M_\pm$ since an observer able to see our system from a large distance 
  	   is situated in a nonrotating frame whereas our rotating ones are bounded by
  	   the light cylinders.} 
  
  Or, as an alternative, we may assume that in this thin but finite layer of matter there is 
  a still thinner (but also finite) intermediate vacuum layer $\M_{\rm int}$. Then nothing 
  prevents us to use one reference frame in $\M_{\rm int}$ for matching it to the matter 
  layers on the ``positive''  side with $\omega > 0$ and another one for matching to 
  the  ``negative'' side with $\omega < 0$ that will in turn smoothly join the vacuum 
  region $\M_-$.        
  
  This reasoning, though apparently correct, still looks rather artificial, and it would be
  more preferable to obtain a model in which both $\Omega(\M_\pm) > 0$. It can be shown,
  however, that at least with the present internal solution \rf{L1}--\rf{L3} such a model 
  cannot be obtained, see the Appendix. We can hope that other sources of gravity in 
  the internal region can provide such a model.

\section*{Appendix}
\sequ{0}
\def\theequation{A.\arabic{equation}}

  Let us try to modify the model built in Section 3 by assuming $E_0 > 0$, so that it could
  be matched to both $\Omega(\M_\pm) > 0$. As before, we assume $\omega_0 > 0$
  and use the notations $y = x/x_0$. For $E(x)$ we can write 
\beq                      \label{A1}
	E = \frac {r_0(1-y^2)} {2} \bigg[ \frac {2y}{1-y^2}	+  \ln\frac{1+y}{1-y} + E_0\bigg]. 	
\eeq    
  The matching conditions on $\Sigma_\pm$ have the same form \rf{ju-1}. The condition 
  $[\mu] =0$ does not affect our consideration. The conditions $[\gamma]=0$ and $[\beta] =0$, 
  as before, lead to 
\bearr                    \label{A2}
		Q^2 x_0^2 (1-y_\pm^2) = 1 - \Omega^2 X_\pm^2 =: P_\pm,
\yyy				\label{A3}
		\frac{r_0^2}{Q^2 x_0^2 (1-y_\pm^2)} = \frac {X_\pm^2}{1 - \Omega^2 X_\pm^2}.  		  
\ear
  From \rf{A2}, \rf{A3} it follows $X_\pm^2 = r_0^2$. Furthermore, the condition $[E]=0$ yields,
  instead of \rf{not-P},
\bearr  			\label{A4}
		P(y) = (1- y^2) \Big[ 1 - (L+ E_0) 
\nnn \inch		
		- \frac {(1 - y^2)}{4} (L+ E_0)^2 \Big],
\ear 
  where we have omitted the index $\pm$ near $P$ and $y$, and, as before,
  $L = \ln\dfrac{1+y}{1-y}$. 
   
  Our task is to find, instead of $y_- = -y_+$, such values of $y_\pm$ for the surfaces $\Sigma_\pm$ 
  that 
\begin{enumerate}  
\item
  	$y_+ > 0$ and $y_- < 0$ (to provide the \wh\ nature of the internal region where $y=0$ is the throat);
\item  
  	Both $E(y_\pm) > 0$, to be matched with $E = \Omega r_0^2$ and to provide $\Omega > 0$ 
  	in $\M_\pm$  	   
\end{enumerate}  
\begin{figure}  
\centering
          \includegraphics[width=7cm]{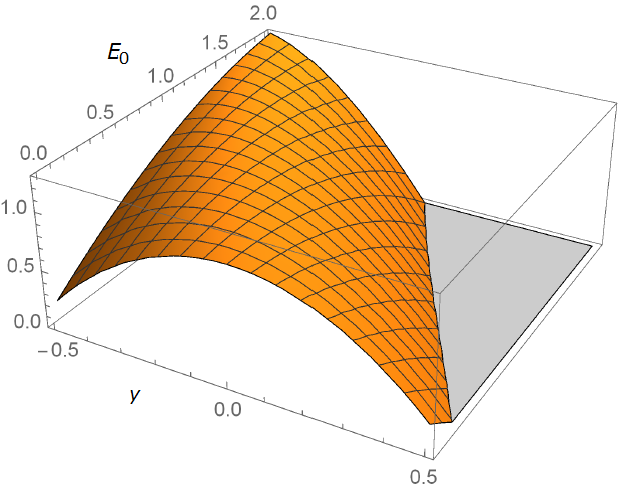}
\caption{\small
          The function $H(y, E_0)$
              }  
\end{figure}  
    
  From \rf{A2} it follows that the function 
\[  
  		H(y, E_0) := P(y)/(1 - y^2)  
\]  
  takes the same value equal to $Q^2 x_0^2$ on both junctions, that is, $H(y_+) = H(y_-)$. As a 
  function of $y$, $H$ has a single $E_0$-dependent maximum (see Fig.\,2), hence $y_+$ and 
  $y_-$ are located on different sides of this maximum. 
  
  On the other hand, $E(y)$ given by \rf{A1} is a monotonically growing function and takes a zero 
  value at the point $y = y_* < 0$ where $L + E_0 = -2y/(1-y^2)$. By Requirement 2, both $y_\pm$ 
  should be located on the $y$ axis to the right of $y_*$. However, it is straightforward to verify that, 
  at any fixed $E_0$, $dH/dy (y_*) < 0$, that is, this point is located on the descending part of the
  plot of $H(y)$, to the right of its maximum, hence only $y_+$ can be larger than  $y_*$, and 
  Requirement 2 cannot be fulfilled. 
  
    We see that the present solution cannot lead to a model with $E(y_\pm) > 0$.

\subsection*{Acknowledgments}

\small
  This publication was supported by the RUDN University program 5-100.
  The work of KB was performed within the framework of the Center FRPP 
  supported by MEPhI Academic Excellence Project (contract No. 02.a03.21.0005,  27.08.2013)
  and by Russian Basic Research Foundation Grant 19-02-00346.   
  The work of VK was supported  by the Ministry of Education 
  and Science of Russia in the framework of State Contract 9.1195.2017.6/7. 


\end{document}